\documentclass[twocolumn]{emulateapj}

\slugcomment{Astrophysical Journal accepted version}

\shorttitle{
Imaging the accretion flow in M87 by space VLBI
}
\shortauthors{Takahashi and Mineshige}
\usepackage[usenames]{color}
\usepackage{graphicx}

\begin{document}

\title{
Constraining the size of the dark region around the M87 black hole by space-VLBI observations
} 

\author{Rohta Takahashi\altaffilmark{1}}
\affil{High Energy Astrophysics Laboratory, the Institute of Physical and Chemical Research, 2-1 Hirosawa, Wako, Saitama 351-0198, Japan}
\email{rohta@riken.jp}

\and

\author{Shin Mineshige\altaffilmark{2}}
\affil{Department of Astronomy, Kyoto University, Sakyo-ku, Kyoto 606-8502, Japan}
\email{shm@kusastro.kyoto-u.ac.jp}


\begin{abstract}
In order to examine if the next generation space VLBI,
such as VSOP-2 (VLBI Space Observatory Programme-2),
will make it possible to obtain direct images of the accretion flow around the M87 black hole, we calculate the expected observed images
by the relativistic ray-tracing simulations under the considerations
of possible observational errors.
We consider various cases of electron temperature profiles,
as well as a variety of the distance, mass, and spin of the M87 black hole.
We find it feasible to detect an asymmetric intensity profile around the black hole
 caused by rapid disk rotation,
as long as the electron temperature does not steeply rises towards the black hole, as was predicted by the accretion disk theory and the three dimensional 
magnetohydrodynamic simulations. 
Further, we can detect a deficit in the observed intensity around the black hole
when the apparent size of the gravitational radius is larger than 
$\gtrsim 1.5 \mu$ arcseconds. 
In the cases that the inner edge of the disk is located at the radius of
the innermost stable circular orbit (ISCO), moreover,
even the black hole spin will be measured. 
 We also estimate the required signal-to-noise ratio $\mathcal{R}_{\rm SN}$ 
for achieving the scientific goals mentioned above,
 finding that it should be at least 10 at 22 GHz.
To conclude, direct mapping observations by the next generation space VLBI
will provide us a unique opportunity 
to provide the best evidence for the presence of a black hole
and to test the accretion disk theory. 
\end{abstract}




\section{Introduction}
Elucidating the nature of space plasmas and their dynamical behavior in a strongly curved 
space-time remains one of the greatest challenges of the observational astronomy 
and astrophysics in this century.  This requires a superb angular resolution, down to tens 
of micro-arcsec and, hence, has been impossible up to now with any instruments 
at any wavelengths. Surprisingly, however, the situations is expected to be enormously 
improved in the near future thanks to the rapid progress of the very long baseline 
interferometer (VLBI) technique from the radio to sub-millimeter band. 
In many past studies, the observational feasibility of the direct imaging of the black hole 
shadow (or the black hole silhouette) with sub-millimeter interferometers on the ground 
has been investigated for the Galactic Center black hole 
Sgr A* \citep[e.g.][]{fma00,m04,m07,f09,d09,hts09}
and the black hole in the elliptical galaxy M87 \citep[][]{bl09}. 
Following the same line, 
in this paper we elucidate the feasibility study of obtaining the direct imaging 
of the black hole with planned space-VLBI satellite. 

It is long believed that radio emission arises from bi-directional jets emanating from 
the accretion flow, and not from the accretion flow itself.  This may not be the case, however, 
if there exist high-energy electrons with energy above 100 keV and if the magnetic field 
strengths are on average greater than about one percent of the equi-partition value. If these 
two conditions are satisfied in accretion flow, we expect significant radio emission not only 
from the jets but also from the flow itself \citep[see, e.g.,][]{nym95,ny95}. It is true that the radio 
emission from the flow is usually overwhelmed by that from jets owing to much higher 
antenna temperature of the jets, but it is nevertheless possible, in principle, to separately 
detect radio emission from the accretion flow if observed with instrumentations with good 
spatial resolution. Future direct imaging observations of the innermost region of AGNs with 
VLBI will provide unique and excellent opportunities to map the accretion flow plunging into 
black holes \citep[e.g.,][]{fma00, m04, t04, t05, bl06a, bl06b, ysh06, nlgb07, hcsy07, h08,bl09,
nt10}. Such observations will also prove the Kerr geometry \citep{b72, t04, t05} predicted 
by the gravity theories, including general relativity \citep{p08a, p08b}. 

One of the leading, on-going projects of the radio VLBI is the VSOP-2 with the ASTRO-G satellite,   
space VLBI program with the highest angular resolution of 38 $\mu$as at the  
observation frequency of 43 GHz \citep[e.g.][]{h05,t08,h09}. 
The minimum antenna temperature which 
can be detected by the VSOP-2/ASTRO-G is about a few times $10^8$ K, which is far exceeded 
by the expected temperature of optically thin accretion flow 
\citep[e.g.][]{nym95,ny95,enoy96,nmkk96,mmk97,yqn03}. 
For details, see the mission web site (\verb"http://www.vsop.isas.ac.jp/vsop2e/"). 
There is another on-going space-VLBI mission called 
RadioAstron (for details, see, \verb"http://www.asc.rssi.ru/radioastron/"). 
We have mainly used the parameters of the space-VLBI satellite in VSOP-2/ASTRO-G 
as one example and the results presented in this paper can be applied to other similar 
space-VLBI missions. 

The existence of a black hole in an accretion flow or a disk will manifest itself as a region 
with a deficit intensity, sometimes called as {\lq\lq}black hole shadow{\rq\rq} or 
{\lq\lq}black hole silhouette{\rq\rq}, in the observed intensity map. 
The properties of the black hole shadow were investigated for optically thick disks 
\citep[e.g.][]{l79,fy88} and for optically thin flow \citep[e.g.][]{fma00}. 
The angular size and precise shape of the black hole shadow depend on the distance, 
black hole mass, space-time geometry, and the optical depth of the accretion flow 
at the observed frequencies \citep[e.g.][]{b72,l79,fy88,t04,t05}.
 
There are good targets of supermassive black holes suitable for the VLBI imaging observations. 
For the observations at 43 GHz, 
the best target will be M87, which has the second largest angular size of the gravitational radius, 
$r_{\rm g}/D \sim$ 1-3 $\mu$as. Here, $r_{\rm g}\equiv GM/c^2$, and $M$ and $D$ are the mass 
and the distance of the black hole, respectively. According to the relativistic ray-tracing calculations 
under the Kerr geometry, the maximum size of the shadow ranges between 4--15 $r_{\rm g}/D$ 
\citep[see, Fig. 2 in][]{t04} which corresponds to 6--24 $\mu$as for M87 when 
$M=3\times10^9M_\odot$ and $D=16$ Mpc are assumed where $M_\odot$ is the solar mass. 
Note that the black hole in the Galactic Center (Sgr A*), the largest black hole in terms of 
the apparent angular size of $r_{\rm g}/D \sim 5~\mu$as, is not suitable for  
the observations at 43 GHz, since the image of the accretion flow around Sgr A* will be totally 
washed out at low frequencies{, $\lesssim 100\sim 150$GHz, }
because of interstellar scattering.  In fact, a rather broadened 
radio morphology with an elliptic shape was already obtained 
{at the observed frequency 86 GHz} 
(e.g., Shen et al. 2005, see also Doeleman et al. 2008 
{for the observations at 230 GHz}). 

In this study, we investigate how the accretion flow around the black hole in M87 will 
be observed and mapped 
by technically feasible space-VLBI observations at 43 GHz and 22GHz. 
For the bolometric luminosity of $L_{\rm bol}\sim 10^{41}$ erg s$^{-1}$ 
\citep{bsh91,rdfhc96,dm03} and the black hole mass of $M\sim 3 \times 10^9$ M$_\odot$, 
the Eddington ratio of M87 is about $L/L_{\rm E} \sim 10^{-6}$, much less than unity, 
indicating that the accretion flow is likely to be radiatively inefficient 
\citep[see Chapter 9 of][]{kfm08}.  
The observed multi-wavelength continuum spectrum of M87 from radio to X-rays can 
be explained in terms of the RIAF (radiatively inefficient accretion flow) model 
(or its variant, advection-dominated accretion flow, ADAF model) \citep{dm03,wlwz08,bl09,l09,nt10}. 
Further, the powerful, relativistic jets are observed in 
M87, which seem to be produced in the innermost part of the accretion flow. 
To understand the rapid TeV $\gamma$-ray variability in M87 jets detected by 
the High Energy Steroscopic System (HESS, Aharonian et al. 2006), it is suggested that 
the black hole in M87 may be rapidly rotating \citep{wlwz08}. 
On this ground, we calculate the expected images of the central region of M87 
by the radiative transfer calculations in the Kerr space-time, basically adopting 
the same method as that used in \cite{t04}, \cite{tw07} and \cite{th10}
except that we employ the RIAF model, instead of the standard disk model, in the present study.  
The continuum emission at around 43 GHz will be of Rayleigh-Jeans spectrum 
because of the self-absorption of the synchrotron radiation by the thermal electrons. 
We, therefore, assume that the accretion flow is optically thick to absorption at 43 GHz. 

The plan of this paper is as follows:  
In \S 2 we explain the methods and models in this work. 
The results of numerical simulations are given in \S3. 
The important issues are discussed in \S 4 and the conclusions are given in \S 5.

\begin{figure}
\begin{center}
\includegraphics[height=8.5cm,width=8.5cm]{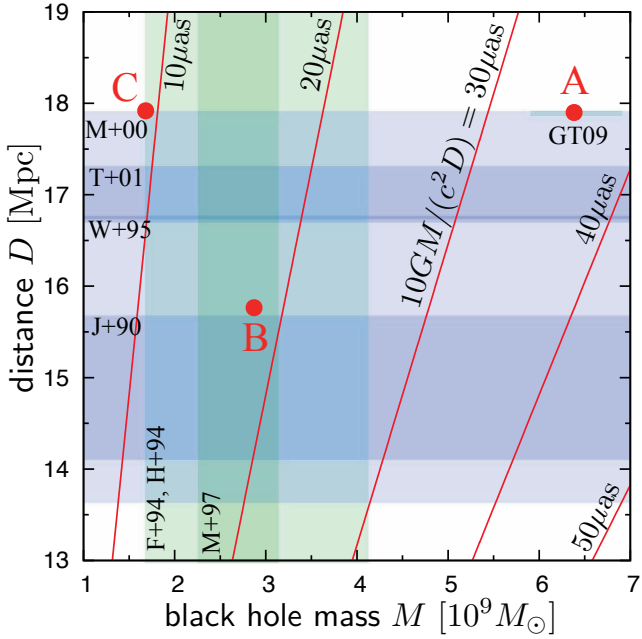}
\end{center}
\caption{\label{fig:Md}
Diagram of the black hole mass, $M$, and the distance, $D$, 
with the observational data points of M87.  
Lines of $10GM/(c^2 D)=10\mu$as, $20\mu$as, 
$30\mu$as, $40\mu$as and  $50\mu$as are also indicated. 
The ranges of the black hole mass of and 
the distance to M87 reported in the past studies are plotted by the shaded region. 
{
For the black hole mass, we use $M=(2.4\pm 0.7)\times 10^9 M_\odot$ 
[\cite{f94}(F+94); \cite{h94}(H+94)], $(3.2\pm 0.9)\times 10^9 M_\odot$ [\cite{m97} (M+97)] 
and $(6.4\pm 0.5)\times 10^9 M_\odot$ [\cite{gt09} (GT09)]. 
The mass in GT09 is calculated by assuming $D=17.9$ Mpc. 
}
For the distance to M87, we use $D=14.7\pm 1.0$ Mpc [\cite{j90}(J+90)], 
$16.75$ Mpc [\cite{w95}(W+95)], $16.0\pm 1.9$ Mpc [\cite{m99}(M+00)] and 
$17.0\pm 0.3$ Mpc [\cite{t01}(T+01)].  
The values of $M$ and $D$ adopted in the present study 
are denoted by A, B and C, respectively  (see also Table \ref{table:Md}). 
}
\end{figure}


\section{Methods and Models}

We first need to specify the black hole mass $M$ and the distance $D$ to M87, 
but there are uncertainties in the observed values of $M$ and $D$. 
In Fig.\ref{fig:Md}, we summarize the past reports for the mass and the distance measurements. 
In this figure, we plot three data for the mass measurements and four data for the distance 
measurements with uncertainties indicated by the shaded region (for the details of the data, 
see the caption of Fig.\ref{fig:Md}). 
In this study, we adopt three datasets of combination of $M$ and 
$D$ denoted as A, B and C in Fig.\ref{fig:Md} and these values are summarized in 
Table \ref{table:Md}. 
In contrast, the black hole spin is not so well constrained by the observations.
Therefore, we consider two extreme cases: a black hole with spin parameter $a=0$ 
(i.e., a Schwarzschild black hole) and one with $a=0.998$ (i.e., an extreme Kerr black hole) 
where $a=J/GM^2c^{-1}$ where $J$ is the angular momentum of the black hole.

\begin{deluxetable*}{cccc}
\tabletypesize{\normalsize}
\tablecaption{
{
The black hole mass of and the distance to M87 adopted in this study. 
The corresponding apparent angular sizes are also listed. 
}
\label{table:Md}}
\tablewidth{0pt}
\startdata
\hline
model & A & B & C\\
\hline
black hole mass $M$ [$10^9M_\odot$] & {6.4} & 3.0 & 1.7 \\
distance to M87 $D$ [Mpc]& {17.9} & 16.0 & 17.9 \\
{
angular size corresponding to 
}
$10 GM/(c^2 D)$ [$\mu$as] & {35.0} & 18.5 & 9.3 
 \enddata
\end{deluxetable*}

\begin{figure*}
\begin{center}
\includegraphics[height=8cm,width=16cm]{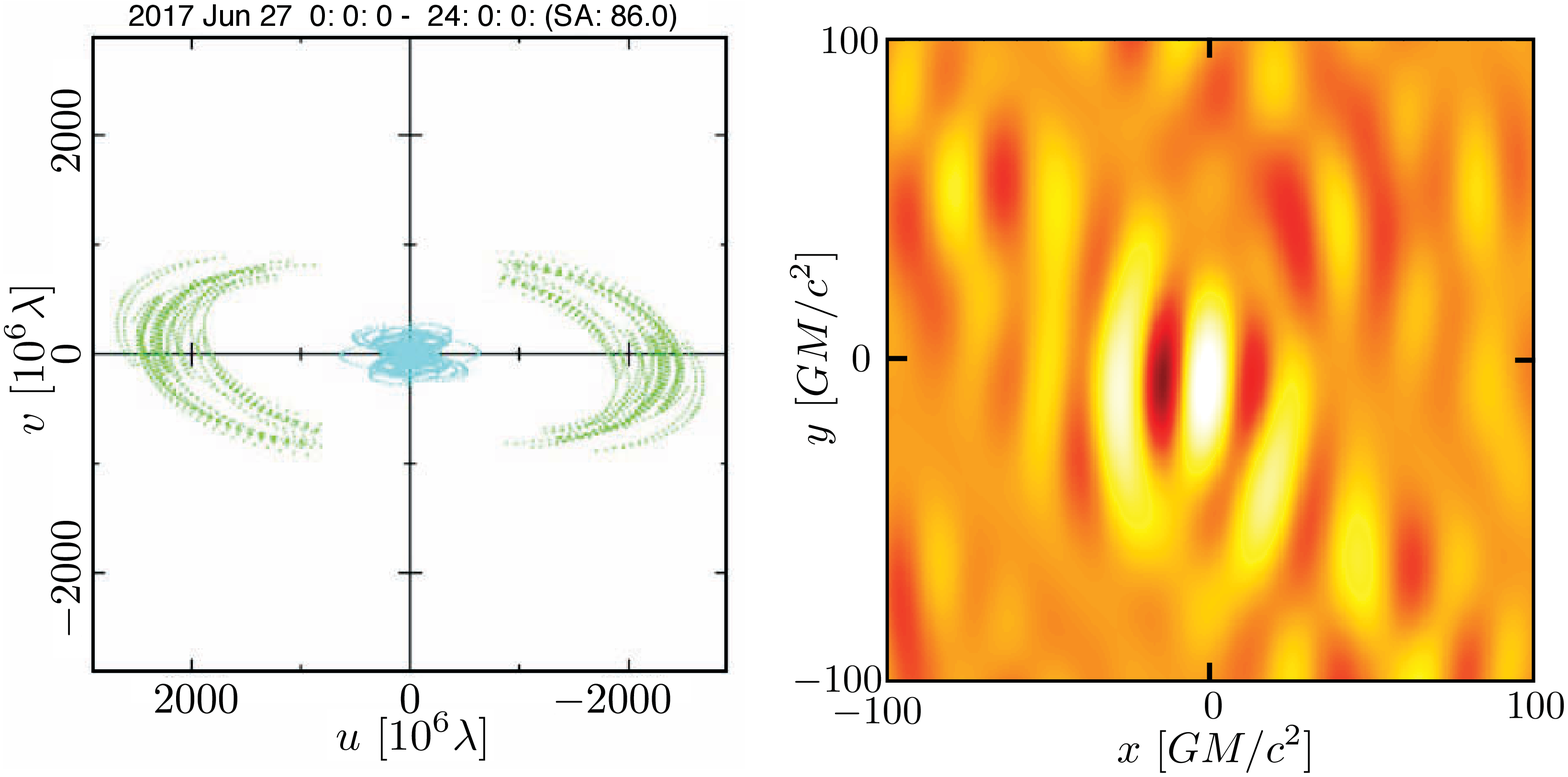}
\end{center}
\caption{\label{fig:beam}
{
A sample of the $u$-$v$ coverage of the VSOP-2 for the observation of M87 within one day ({\it left}) 
and the corresponding synthesized dirty beam for model B ({\it right}). 
}
}
\end{figure*}

Next, we need to prescribe electron temperature distribution. According to the RIAF 
model, ion temperatures are very high, 
close to the virial, and are several times $10^{12} (r/r_{\rm g})^{-1}$ K near the black hole 
where $r$ is the radial distance from the center, 
whereas the electron temperature will be much less and saturated around a few times 
10$^9$K (which is about $m_{\rm e}c^2/k$ with $m_{\rm e}$ and $k$ being the electron 
mass and the Boltzmann constant, respectively) due to the enhanced cooling in the 
relativistic regimes \citep[e.g.,][see Kato, Fukue, \& Mineshige 2008 for a review and references therein]
{ny95}.  
Note that the accretion disk corona model also gives similar electron temperature profiles 
but with slightly lower values \citep[e.g.,][]{lmo03}. As a result, the electron temperature is 
spatially uniform close to the black hole while the ion temperature increases in the central 
region of the disk. Such features are confirmed by three-dimensional MHD simulation 
\citep[e.g.,][]{okm05}. The expected electron temperature of $> 10^9$ K in the central part of the 
flow certainly exceeds the detection limit of VSOP-2/ASTRO-G at 22 GHz and 43 GHz. 
Hence, we prescribe electron temperature profile as
$ T_{\rm e}=5 \times 10^9 (r/r_{\rm b})^{-p}$ K where $p$ and $r_b$ are parameters. 
In this study, we calculate three cases of $p$: $p=0.1$, 0.5 and 1.0. 
Note that the last model has the same temperature profile as that of ions. 
As for the parameter $r_b$, we determine the value so as to reproduce the observed 
energy spectrum, finding $r_{\rm b}\sim 100 r_{\rm g}$. 
Similar electron temperature 
profiles were used in the past study \cite[e.g.][]{bl09,nt10} and successfully reproduced the observed 
energy spectrum and images. At 22 GHz and 43 GHz, 
the energy spectrum is well explained by thermal synchrotron emission 
\citep[e.g.][]{dm03,wlwz08,l09,bl09,nt10}. 

When the radiative cooling is not so efficient above $\sim 10^{10}$ K 
and/or the energy transportation from ions 
to electrons is very efficient, electron temperatures can increase with decreasing $r$. 
So, the radial profile of the electron temperature contains the important information 
about the electron heating mechanism of the accretion flow. In this study, we show that 
the observed images obtained by the space-VLBI observations
can put useful constraints on the electron temperature profile.

Further, we need to prescribe the inner edge and the velocity pattern of the accretion disk. 
We consider two cases for the inner edge of the luminous part of the accretion disk: one is at the 
innermost stable circular orbit (ISCO), $r_{\rm ISCO}$, and 
the other is at the event horizon, $r_{\rm H}$. 
In this study, we assume the sub-Keplerian velocity profile with the angular velocity given as 
$\Omega=\Omega_{\rm K}/\sqrt{2}$ and the radial component of the four velocity given as 
$u^r=u^r_{\rm FF}/\sqrt{2}$ where $\Omega_{\rm K}$ and $u^r_{\rm FF}$ are the Keplerian 
angular velocity and the radial four velocity component of the freely falling motion with zero angular 
momentum at infinity, respectively. 
We expect that the accretion disk in M87 has the sub-Keplerian velocity distribution near 
the black hole. 
As for the cases of different velocity profiles (such as freely falling motion 
and Keplarian rotation), see discussion in \S4. 
The viewing angle, $i$, between the direction of the observer and the rotation axis of the disk 
is assumed to be $i=45^\circ$. Although in the past studies the inclination angle of $i=30^\circ$ 
is used \citep[e.g.,][]{nt10}, we have confirmed that the main results in this study do not change 
even if $i=30^\circ$ is used. 
Based on the assumptions described above, we perform the general relativistic ray-tracing 
calculations in the Kerr space time;  that is, we fully incorporate the relativistic effects, such as the 
frame-dragging, the gravitational redshift, the bending of light and the Doppler boosting. 
Our calculation methods are prescribed in our past papers, i.e. \cite{tw07}, \cite{nt10} and 
Appendix of \cite{th10}.

\begin{deluxetable*}{ccccccccccc}
\tabletypesize{\normalsize}
\tablecaption{Models \label{table:models}}
\tablewidth{0pt}
\startdata
 & model 	& $M$\&$D$ 	& $a$ & {$i$} & $r_{\rm in}$ 
& $r_{\rm out}$ \tablenotemark{a}
& $\ell_{\rm BH}$ \tablenotemark{a}
& {Asymmetry}
& {Deficit} 
& Fig\\
\hline
$p=0.5$ 	& A(a)p05 	&A	& 0 		&{$45^\circ$}	& $r_{\rm ISCO}$ 	& 66 		&{20}		&{Yes}	&{Yes}	& {3, 4} \\
		& A(b)p05 	&A	& 0 		&{$45^\circ$}	& $r_{\rm H}$ 		& 66 		&{20}		&{Yes}	&{Yes}	& {3, 4, 7} \\
		& A(c)p05 	&A	& 0.998 	&{$45^\circ$}	& $r_{\rm H}$ 		& 66 		&{20}		&{Yes}	&{Yes}	& {3, 4} \\
		& B(a)p05 	&B	& 0 		&{$45^\circ$}	& $r_{\rm ISCO}$ 	& 110 	&26					&{Yes}	&{Yes}	& {3, 4} \\
		& B(b)p05 	&B	& 0 		&{$45^\circ$}	& $r_{\rm H}$ 		& 110 	&24					&{Yes}	&{Yes}	& {3, 4, 5} \\
		& B(c)p05 	&B	& 0.998 	&{$45^\circ$}	& $r_{\rm H}$ 		& 110 	&25					&{Yes}	&{Yes}	&  {3, 4}\\
		& C(a)p05 	&C	& 0 		&{$45^\circ$}	& $r_{\rm ISCO}$ 	& 220 	&38					&{Yes}	&{Yes}	&  {3, 4}\\
		& C(b)p05 	&C	& 0 		&{$45^\circ$}	& $r_{\rm H}$ 		& 220 	&39\tablenotemark{b}	&{Yes}	&{No}	& {3, 4} \\
		& C(c)p05 	&C	& 0.998 	&{$45^\circ$}	& $r_{\rm H}$ 		& 220 	&38\tablenotemark{b}	&{Yes}	&{No}	& {3, 4}\\
\hline
$p=0.1$ 	& A(a)p01 	&A	& 0 		&{$45^\circ$}	& $r_{\rm ISCO}$ 	& 66 		&90					&{Yes}	&{Yes}	&  {4} \\
		& A(b)p01 	&A	& 0 		&{$45^\circ$}	& $r_{\rm H}$ 		& 66 		&90					&{Yes}	&{Yes}	&  {4}\\
		& A(c)p01 	&A	& 0.998 	&{$45^\circ$}	& $r_{\rm H}$ 		& 66 		&90					&{Yes}	&{Yes}	&  {4}\\
		& B(a)p01 	&B	& 0 		&{$45^\circ$}	& $r_{\rm ISCO}$ 	& 110 	&110				&{Yes}	&{Yes}	&  {4}\\
		& B(b)p01 	&B	& 0 		&{$45^\circ$}	& $r_{\rm H}$ 		& 110 	&110				&{Yes}	&{Yes}	&  {4}\\
		& B(c)p01	&B	& 0.998 	&{$45^\circ$}	& $r_{\rm H}$ 		& 110 	&110				&{Yes}	&{Yes}	&  {4}\\
		& C(a)p01 	&C	& 0 		&{$45^\circ$}	& $r_{\rm ISCO}$ 	& 220 	&130				&{Yes}	&{Yes}	&  {4}\\
		& C(b)p01 	&C	& 0 		&{$45^\circ$}	& $r_{\rm H}$ 		& 220 	&130				&{Yes}	&{Yes}	&  {4}\\
		& C(c)p01 	&C	& 0.998 	&{$45^\circ$}	& $r_{\rm H}$ 		& 220 	&130				&{Yes}	&{Yes}	&  {4}\\
\hline
$p=1.0$	& A(a)p1 	&A	& 0 		&{$45^\circ$}	& $r_{\rm ISCO}$ 	& 66 		&{16}		&{Yes}	&{Yes}	&  {4} \\
		& A(b)p1 	&A	& 0 		&{$45^\circ$}	& $r_{\rm H}$ 		& 66 		&{15}		&{Yes}	&{Yes}	&  {4} \\
		& A(c)p1 	&A	& 0.998 	&{$45^\circ$}	& $r_{\rm H}$ 		& 66 		&{15}		&{Yes}	&{Yes}	&  {4}\\
		& B(a)p1 	&B	& 0 		&{$45^\circ$}	& $r_{\rm ISCO}$ 	& 110 	&19					&{Yes}	&{Yes}	&  {4}\\
		& B(b)p1 	&B	& 0 		&{$45^\circ$}	& $r_{\rm H}$ 		& 110 	&18\tablenotemark{b}	&{Yes}	&{No}	&  {4} \\
		& B(c)p1 	&B	& 0.998 	&{$45^\circ$}	& $r_{\rm H}$ 		& 110 	&17\tablenotemark{b}	&{Yes}	&{No}	&  {4} \\
		& C(a)p1 	&C	& 0 		&{$45^\circ$}	& $r_{\rm ISCO}$ 	& 220 	&\nodata				&{Yes}	&{No}	&  {4}\\
		& C(b)p1 	&C	& 0 		&{$45^\circ$}	& $r_{\rm H}$ 		& 220 	&\nodata				&{marginal}	&{No}	&  {4}\\
		& C(c)p1 	&C	& 0.998 	&{$45^\circ$}	& $r_{\rm H}$ 		& 220 	&\nodata				&{marginal}	&{No}	&  {4}\\
 \hline
$p=0.5$ 	& {B(a)p05i15} 	& {B}	&  {0}		&{$15^\circ$}	&  {$r_{\rm ISCO}$} 	&  {110} 	& {26}	&{Yes}	&{Yes}	&  {6}\\
		& {B(b)p05i15} 	& {B}	&  {0} 		&{$15^\circ$}	&  {$r_{\rm H}$} 		&  {110} 	& {24}	&{Yes}	&{Yes}	&  {6} \\
		& {B(c)p05i15} 	& {B}	&  {0.998} 	&{$15^\circ$}	&  {$r_{\rm H}$} 		&  {110} 	& {25}	&{Yes}	&{Yes}	&  {6} 
\enddata
\tablenotetext{a}{In units of $r_{\rm g}\equiv GM/c^2$.}
\tablenotetext{b}{Marginal value.}
\end{deluxetable*}

After calculating the simulation images of the intensity $I_\nu$ with observed frequency at  
22 GHz and 43 GHz by assuming the thermal synchrotron spectrum,  
we calculate the expected images smeared out by the spatial resolution of 
$\sim 38 \mu$ arcseconds
with the realistic $u$-$v$ coverage data created by the simulation tool, the Astronomical Radio 
Interferometer Simulator \citep[ARIS;][]{a07}. 
\footnote{
We thank Y. Asaki for providing ARIS. 
}
{
In the left panel of Fig \ref{fig:beam}, we show a sample of the $u$-$v$ coverage 
of the VSOP-2 (ASTRO-G satellite and VLBA) for the observation of M87 within one day. 
By using this $u$-$v$ coverage, 
the synthesized dirty beam is calculated as shown in the right panel of Fig \ref{fig:beam}. 
The scale of the dirty beam shown in this figure is calculated based on model B.  
}

{
The calculation method of the observed image smeared by the spatial resolution of the 
VSOP-2/ASTRO-G observations is as follows: 
1) 
From the dirty beam, the primary lobes of the clean beam can be calculated by fitting  
the primary component of the dirty beam by the elliptical gaussian. 
2) 
We calculate the visibility $V(u,v)$, that is, 
the Fourier components of the elliptic gaussian beam $S(u,v)$ and the theoretical images. 
3) 
The map of the spatially smoothed intensity 
$I_\nu$ is calculated as 
$I_\nu(x,y)=\int \int e^{2\pi i(xu+yv)} S(u,v)V(u,v)du dv$ \citep[e.g.][hereafter, TMS]{tms01}. 
The image produced based on this procedure basically corresponds to the image produced 
by the CLEAN deconvolution algorithm if the algorithm works well and if the effects of the side lobe 
can be completely removed, i.e. no deconvolution errors 
(see, Sec 4 for  important discussion on the data deconvolution method for the image 
containing the dark region around the black hole).    
In the CLEAN deconvolution algorithm (e.g., Sec 11.2 of TMS) 
which is usually used in the radio data analysis, 
the clean map is calculated by first extracting the dirty beam components with some factor 
from the dirty map and then replacing the dirty beam with a gaussian
(or similar functions that are free from negative values). 
If this procedure works well, 
the effects of the side-lobe patterns can be removed and the resultant images 
made by the CLEAN deconvolution algorithm corresponds to the image which is obtained from 
the original image smeared with the gaussian.  
Therefore, the image produced by the calculation method used in this study (as denoted above) 
corresponds to the images with no deconvolution errors and no effects of the sidelobe 
in the CLEAN deconvolution algorithm. About the possible problems in the application of the 
CLEAN deconvolution algorithm to the images with the dark region around the black hole, 
see the discussions in Sec 4. 
}

{
In this study, we calculate the smeared images for 
30 models (for model parameters see Table \ref{table:models}).
We first prescribe the outer radius of (the luminous part of) 
the accretion disk as follows. 
According to the past observed images of M87 by the radio interferometers
the size of an emitting region is about a few to several hundreds of micro arcseconds 
\citep[e.g.][]{j99,d06}.
Note, however, that the actual length scale should depend on the distance, $D$,
for a fixed value of the angular size,
and the gravitational radius depends on the mass, $M$. 
In order that the emitting region size should be
consistent with the past observational data,
we adopt $r_{\rm out}=$ 66, 110, and 220 ($r_{\rm g}$) 
for Models A, B, and C, respectively (see Table \ref{table:models}).
As to the inner radius and the spin parameter, 
we consider three sets of their combinations: 
(a) $(a, r_{\rm in}$)= (0, $r_{\rm ISCO}$), 
(b) (0, $r_{\rm H}$), and 
(c) (0.998, $r_{\rm H}$). 
From the model names, we can get the information of the assumed parameters,
which include
$M$, $D$ (from A, B or C, see Table \ref{table:Md}), 
 $a$, $r_{\rm in}$ [from (a), (b) or (c)],
 and $p$ (from p01, p05, p1). 
}

\begin{figure*}
\hspace*{-0.5cm}
\begin{center}
\includegraphics[height=11.5cm,width=17cm]{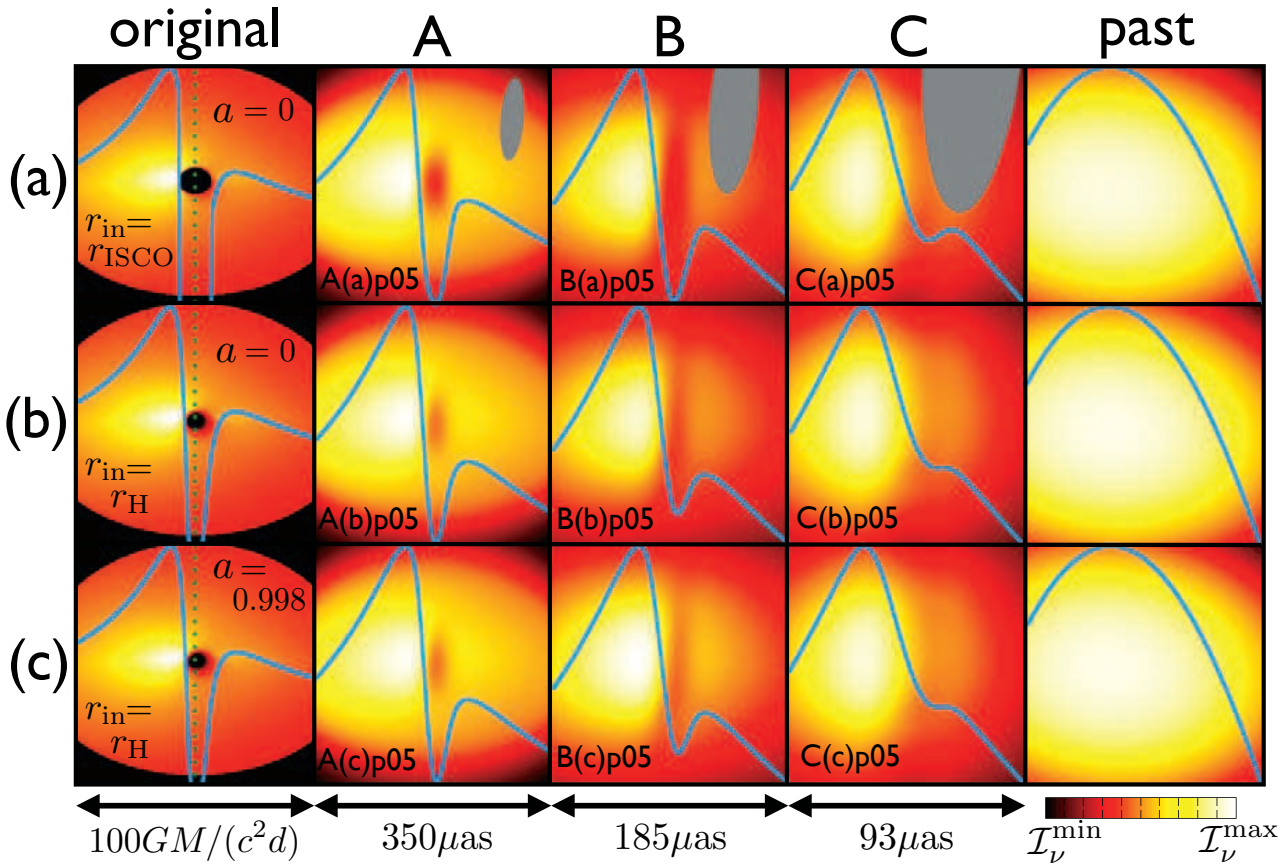}
\end{center}
\caption{\label{fig:image} 
{
Images of an accretion flow surrounding a black hole with the characteristics of M87. 
The emitting accretion flow is assumed to have a sub-Keplarian velocity field with 
an electron temperature profile of $T_e=5\times 10^9$K $(r/r_{\rm b})^{-p}$ 
with $p=0.5$ and $r_{\rm b}=100~r_{\rm g}$. From the left to right panels,
the theoretical (unsmeared) images ({\it first column}), 
the images smeared with the $u$-$v$ coverage of the 
next-generation space-VLBI observations 
[with the beam size of $38\mu$as$\times 120\mu$as ({\it gray ellise})]
for models A, B and C (in Table \ref{table:Md}) 
({\it second, third and forth columns}), 
and the images smeared by the beam size of the past observation 
(with 0.3 mas$\times$0.2 mas, \citep{d06} ({\it fifth column}), respectively. 
The black hole is either (a, b) non-rotating ($a=0$) or (c) maximally rotating ($a=0.998$).
The viewing angle is fixed to be $i=45^\circ$ in this figure. 
The inner edge of the accretion flow is either at (a) the radius of 
the innermost stable circular orbit (ISCO) $r_{\rm ISCO}$ or (b, c) 
at the event horizon $r_{\rm H}$. The left panels show the results based on models 
A(a)p05, A(b)p05, and A(c)p05,
while the right panels are based on models B(a)p05, B(b)p05, and B(c)p05. 
These model parameters are summarized in Table \ref{table:models}. 
The rotation axis of the accretion flow and the black hole is
also shown ({\it green dotted lines}) in the left panels. 
The normalized intensity variation along the horizontal line passing 
the black hole center are overlayed ({\it blue lines}). 
The vertical and horizontal width 
of images is 100 $GM/(c^2 D)$ which corresponds to about 295$\mu$as (A), 
185$\mu$as (B), and 93$\mu$as (C). 
}
}
\end{figure*}

\begin{figure*}
\hspace*{-0.5cm}
\begin{center}
\includegraphics[height=11.5cm,width=17cm]{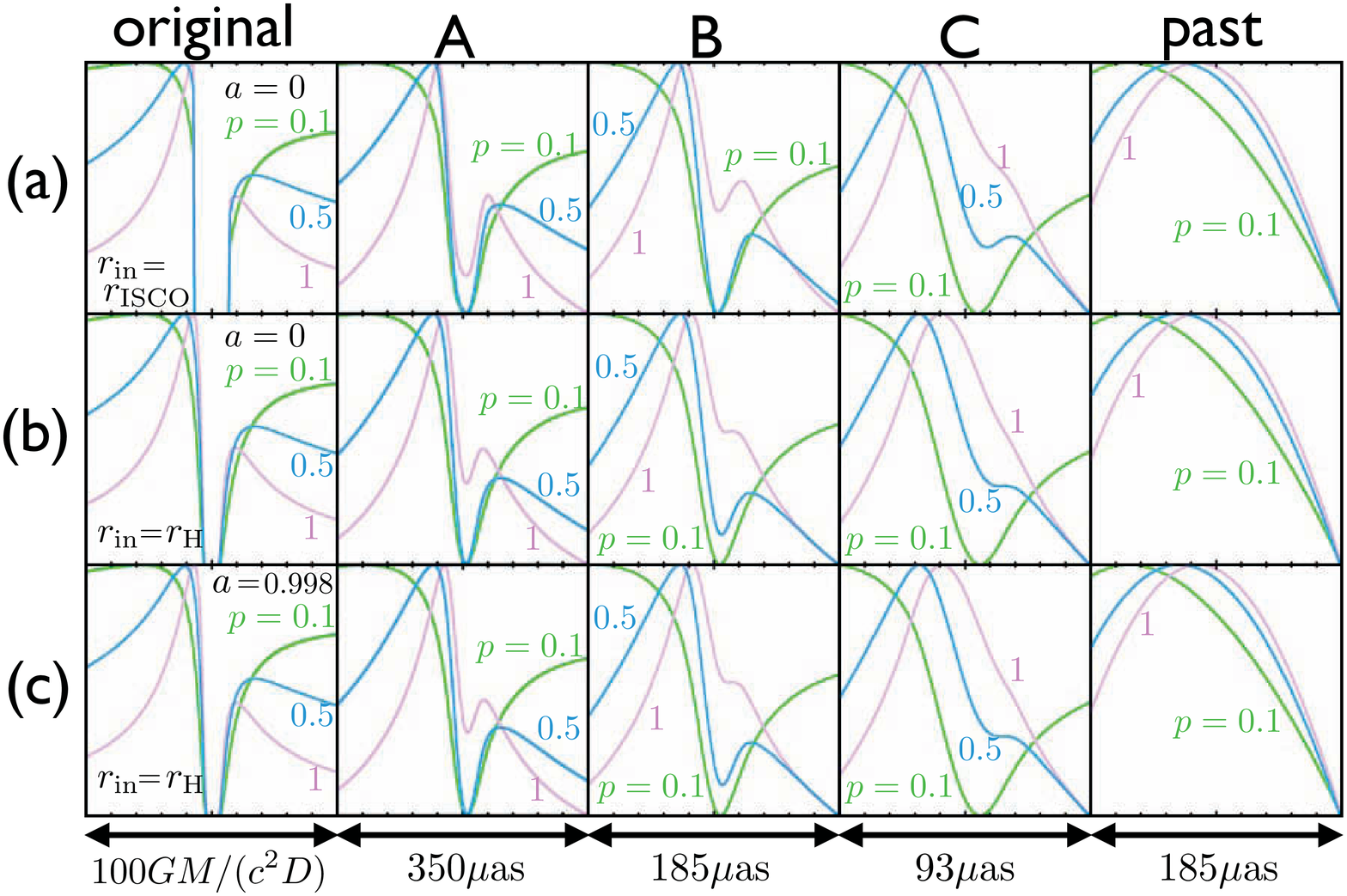}
\end{center}
\caption{\label{fig:cs}
{
The normalized intensity variations along the line passing the black hole center
 for $p=0.1$ ({\it green lines}), 0.5 ({\it blue lines}),
 and 1 ({\it pink lines}),
respectively.  All the lines for 
$p=0.5$ are the same as those in Fig.\ref{fig:image}. 
The normalized intensity variations are shown 
for the theoretical (unsmeared) images ({\it first column}), 
the images smeared with the $u$-$v$ coverage of the next-generation
space-VLBI observations for models A, B and C (in Table \ref{table:Md}) 
({\it second, third and forth columns}), 
and the images smeared with the
resolution of the past observation resolution, respectively
The lines of $p=0.5$ are the same as those in
 Fig. \ref{fig:image} ({\it fifth column}). 
The vertical width is 
100 $GM/(c^2 D)$ which corresponds to about 295$\mu$as (A), 185$\mu$as (B),
and 93$\mu$as (C). 
See Table \ref{table:models} for model parameters of the plotted 27 models.
}
}
\end{figure*}

\section{Appearance of the accretion disk around a black hole}

\subsection{Overview}
Fig.\ref{fig:image} shows the results of the observed intensity distributions for the case of $p=0.5$ 
(i.e., $T_{\rm e}\propto r^{-0.5}$). 
From the left to right panels, we show the theoretical  
(unsmoothed) images ({\it first column}), images smeared with the $u$-$v$ coverage 
of the space-VLBI observations based on models A, B and C 
in Table \ref{table:Md} ({\it second, third and forth columns}), and the images smeared with an elliptical 
beam size of 0.3 mas$\times$0.2 mas as  {in the past observations \citep{d06}} ({\it fifth column}). 
{
From the top to bottom panels, we show the results for the above cases 
(a) ({\it top}), (b) ({\it middle}) and (c) ({\it bottom}). 
}
In this figure, 
the normalized intensity variation along the horizontal line
passing the center is also shown for each image ({\it blue solid line}). 
For other informations, see the caption of Fig.\ref{fig:image}. 

For all the cases of the space-VLBI observations 
({\it second, third and forth columns}), 
the asymmetry of the observed intensity with respect to the vertical line passing the intensity peak 
can be clearly seen, while in the smeared images by the resolution of the past observation 
({\it fifth column}) such asymmetry can not be seen. 
In all the cases of the space-VLBI observations except models of C(b)p05 and C(c)p05, 
furthermore, an under-luminous region (i.e. deficit 
of the observed intensity) around the central black hole can be seen; that is, second 
peak of the intensity can be clearly seen. 
The size of the under-luminous region is nearly the same as that of the beam size ({\it gray ellipse})
for models of A(a)p05, A(b)p05, A(c)p05 and B(a)p05 and is smaller than the beam size for 
models B(b)p05, B(c)p05 and C(a)p05. On the other hand, the intensity contrast amounts to 
$\sim 40$\% for A(a)p05, $\sim 30$\% for A(b)p05, A(c)p05, and B(a)p05, 
$\sim 20$\% for models B(b)p05 and B(c)p05, and $\sim 10$\% for model C(a)p05. 
In terms of the black hole spin, we can not see the difference between $a=0$ and $0.998$, 
in the smeared images of the cases of $r_{\rm in}=r_{\rm H}$ 
(compare {\it middle and bottom}), 
while we can see the difference with $\sim 10$\% difference of the intensity  
in the case of $r_{\rm in}=r_{\rm ISCO}$ (compare {\it top and bottom}).  
Here, note $r_{\rm H}\sim r_{\rm ISCO}$ for $a=0.998$. 

Here, we define the size $\ell_{\rm BH}$ as the length between the two peaks 
of the intensity profile in units of $GM/c^2$, which gives the size of the under-luminous region 
around the central black hole, i.e. from the value of $\ell_{\rm BH}$, 
the spatial size containing the black hole can be measured. 
When there is only one peak in the intensity profile, 
the size $\ell_{\rm BH}$ can not be defined. In Table \ref{table:models}, we summarize the values 
of $\ell_{\rm BH}$ for all models. 
In the case of Sgr A*, the VLBI observations put constraints on 
the size of the region containing the black hole as $\sim 25.2~GM/c^2$ from 
the size measurements of the observed intensity \citep[][]{sllhz05}.  
In Table \ref{table:models}, we can see that models 
A(a)p05, A(b)p05, A(c)p05, B(a)p05, B(b)p05 and B(c)p05 give constraints of $\ell_{\rm BH}$ 
around $\ell_{\rm BH}\sim 25~GM/c^2$ corresponding the size obtained in Sgr A*. 
In all models with $p=0.1$, we find $\ell_{\rm BH}\gtrsim 100$. 
In models of A(a)p1, A(b)p1, A(c)p1, B(a)p1, B(b)p1 and B(c)p1, on the other hand, 
the size $\ell_{\rm BH}$ is smaller than $\lesssim 19~GM/c^2$ 
which gives a more stringent constraint than the case of Sgr A*. 
If this is the case, the space-VLBI observations can put useful constraints on 
the size $\ell_{\rm BH}$ of the region containing the black hole. 
{
In Table \ref{table:models}, we also indicate whether the asymmetry ({\it ninth column})
and the deficit ({\it tenth column}) in the intensity profile can be seen or not 
for all the calculated models. 
Although the asymmetric signature can be seen (or marginally seen) for all the models, 
the deficit can not be seen for some models. 
}

\subsection{Temperature profiles}
Next, Fig \ref{fig:cs} compares the cases of different electron temperature profiles: $p=0.1$ 
({\it green solid line}), 0.5 ({\it blue solid line}) and 1 ({\it pink solid line}). In this figure, we show 
the normalized intensity variation along the line passing the center for 27 models listed in 
Table \ref{table:models}. 
All the lines for $p=0.5$ are the same as those of Fig \ref{fig:image} and the lines 
for $p=0.1$ and 1 are calculated in the same way as those in Fig \ref{fig:image}. 
For all the cases of the space-VLBI observations
({\it second, third and forth columns}) except for models of C(b)p1 and C(c)p1, 
the asymmetry of the observed intensity with respect to the vertical line passing the intensity peak 
can be clearly seen, while in the cases of C(b)p1 and C(c)p1, 
only the asymmetry can be marginally seen. 

On the other hand, we can clearly see the under-luminous region around the black hole 
for all models except C(b)p05, C(c)p05, B(b)p1,  B(c)p1,  C(a)p1,  
C(b)p1 and C(c)p1. 
In general, the steeper the radial temperature profile is (e.g. $p=1$), the more difficult it 
becomes to see 
the asymmetry of the intensity and the under-luminous region around the black hole. 
This is because the size of the effectively emitting region for $p=1$ 
is smaller, compared with other cases ($p=0.1$ and 0.5) and so the luminous part of the 
image tends to be smeared out. 
The observational feasibility of the effects of the black hole spin is basically 
the same as that in the case of $p=0.5$ described above, i.e. 
in the smeared images of the cases with $r_{\rm in}=r_{\rm H}$ (compare {\it middle and bottom}), 
we cannot see a difference between the cases with $a=0$ and $0.998$, 
while in the case with $r_{\rm in}=r_{\rm ISCO}$ (compare {\it top and bottom}) 
we can see. 

\subsection{Observational errors}
So far, we implicitly neglected the effects of the noise introduced by the observations. 
Even when the wide $u$-$v$ coverage is achieved by the space-VLBI satellite, 
the observational features investigated above if noise is large. 
will not be completely obtained. The noise is caused by several factors;  
e.g. deviations of the antenna surface from the ideal profile (for details, see, TMS). 
The complex visibility $V$ is deviated by the noise as $Z=V+\varepsilon$ where 
$Z$ is the measured visibility (including the effects of the noise) and $\varepsilon$ is the noise 
components (see, Fig 6.8 in TMS). The effects of the noise can be included by assuming 
Gaussian noise of standard deviation $\sigma_{\rm noise}$,  
and the probability distributions of the measured visibility is given in Secs 6.2 and 9.3 of TMS. 
In the space-VLBI observations, the standard deviation $\sigma_{\rm noise}$ is calculated 
as the flux density calculated by using the SEFD $S_{\rm E}$
of the space-VLBI satellite ($S_{\rm E}^{\rm space}$) and 
the ground-based interferometer ($S_{\rm E}^{\rm ground}$) as 
$\sigma_{\rm noise}=[S_{\rm E}^{\rm space}S_{\rm E}^{\rm ground}/(2\Delta\nu \tau_a)]^{1/2}/\eta_{\rm Q}$ where $\Delta\nu$ is the bandwidth, $\tau_a$ is the data averaging time and 
$\eta_{\rm Q}$ is the efficiency factor (see, e.g. Sec. 6.2 of TMS). 
In this paper, we use the signal-to-noise (SN) ratio $\mathcal{R}_{\rm SN}$ defined as 
$\mathcal{R}_{\rm SN}\equiv \bar{\sigma}_{\rm source}/\sigma_{\rm noise}$ where 
$\bar{\sigma}_{\rm source}$ is the spatial average value of the source flux density 
{
in the region within the outer disk radius $r_{\rm out}$. 
}

Based on the probability distribution of the noise (in Sec 9.3 of TMS) and the assumed 
value of $\mathcal{R}_{\rm SN}$, the effects of the noise can be included. 
{
In this paper, we take the following procedure to include the noise effects: 
\begin{enumerate}
\item 
First, we calculate the average flux density $\bar{\sigma}_{\rm source}$ for each model in Table 2. 
\item 
Next, we prescribe a single value of $\mathcal{R}_{\rm SN}$, e.g. $\mathcal{R}_{\rm SN}=50$, and 
calculate the standard deviation of the noise 
$\sigma_{\rm noise}(=\bar{\sigma}_{\rm source}/\mathcal{R}_{\rm SN})$ 
based on the calculated value of $\bar{\sigma}_{\rm source}$ and the assumed 
$\mathcal{R}_{\rm SN}$. 
\item 
Here, we injected a single noise value and input the noise effects into $(u, v)$-points. 
For the single value of $\sigma_{\rm noise}$ calculated in the last stage 
and the gaussian probability distribution function 
of the noise (given in Secs 6.2 and 9.3 of TMS), we can produce random gaussian noises. 
In this stage, 
we randomly input the gaussian noise on each $(u, v)$-point of the measured visibility 
which is calculated from the theoretical image by the Fourier transformation, i.e. 
we have the noisy visibility map in this stage. 
It is noted that at this stage 
since the noise is randomly produced for each visibility point, 
the noise values are different at different $(u,v)$-points.  
\item 
From the noisy visibility calculated at stage 3, 
we convolve the sampling function calculated from the $u$-$v$ coverage of the space-VLBI 
satellite and the ground-based interferometers. 
We then calculate noisy image by the inverse Fourier transformation. 
\item 
We make ten independent noisy images in this stage, i.e. stages 3 and 4 are repeated for  
ten times based on the different gaussian random noise patterns.  
\item 
Finally, we average the ten noisy images. 
\end{enumerate}
In this procedure, we calculate the noisy image from the theoretical original image 
for an assumed value of $\mathcal{R}_{\rm SN}$. 
It is noted that the value of $\mathcal{R}_{\rm SN}$ represents the SN ratio for the 
averaged value of the source flux density (from the definition of $\mathcal{R}_{\rm SN}$) 
and the actual SN ratios at different points of the image (or the visibility) are different. 
In general, 
around the most luminous part of the image where the value of the flux is larger then the averaged 
flux, the SN ratios are much higher than $\mathcal{R}_{\rm SN}$ which represents 
the average SN ratio. 
}

\begin{figure*}
\hspace{-5mm}
\begin{center}
\includegraphics[height=8.5cm,width=16.5cm]{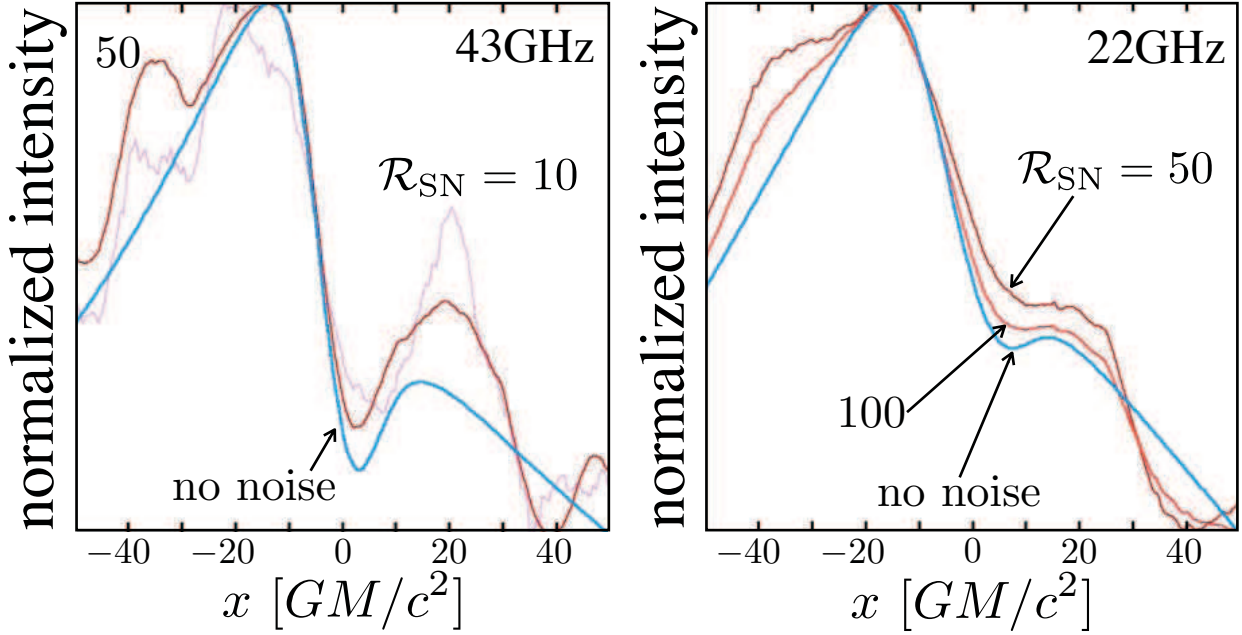}
\end{center}
\caption{\label{fig:noise}
Normalized intensity variations along the line passing the black hole center for 
model B(b)p05 without noise ({\it the blue lines}) and with noise ({\it the brown lines}) 
of the signal-to-noise ratio 
$\mathcal{R}_{\rm SN}=10$ and 50 for the observed frequency 43 GHz ({\it left panel}) and 
$\mathcal{R}_{\rm SN}=50$ and 100 for 22 GHz ({\it right panel}).  
For the case of 43GHz, the curve of no noise is same as the curves of the model B(b)p05 shown in 
Figs \ref{fig:image} and \ref{fig:cs}. 
}
\end{figure*}

In the left panel of Fig \ref{fig:noise}, we show the normalized intensity 
variations for model B(b)p05 with noise of $\mathcal{R}_{\rm SN}=10$, and 50 and without noise. 
These curves with noise are calculated by the average of the ten curves with different 
noise patterns. 
In this way, we find that 
a deficit of the intensity can be measured for $\mathcal{R}_{\rm SN}\gtrsim 10$. 
In the curves of $\mathcal{R}_{\rm SN}=10$ and 50, another deficit of the intensity 
caused by the noise can be seen around $x=40$. 
The depths of these deficits (around $x\sim 40$) are comparable to those of the deficit 
in the center caused by the existence of the black hole. 
We also calculated the case for $\mathcal{R}_{\rm SN}=100$ and confirmed that 
the deficits around $x=40$ disappear. 

Here, for the ASTRO-G satellite, 
we estimate the value of $S_{\rm E}^{\rm space}$ satisfying the condition of 
$\mathcal{R}_{\rm SN} \gtrsim 10$ or $\gtrsim100$. 
By using the values of $\bar{\sigma}_{\rm source}=1$ Jy for M87, 
$S_{\rm E}^{\rm ground}=1500$ Jy for the Very Large Baseline Array (VLBA), 
$\eta_{\rm Q}=0.7$, $\Delta\nu=256$ MHz and $\tau_a=60$ s, 
the SEFD required for 
the observations satisfying $\mathcal{R}_{\rm SN} \gtrsim 10$ 
is $S_{\rm E}^{\rm satellite} \lesssim 1.0\times 10^5$Jy, and  
for $\mathcal{R}_{\rm SN} \gtrsim 100$, the required SEFD is 
$S_{\rm E}^{\rm satellite} \lesssim 1.0\times 10^3$Jy. 
{
At 43 GHz, 
the nominal and the possible worst values of $S_{\rm E}^{\rm satellite}$ 
are 28000 Jy and 190000 Jy for the ASTRO-G satellite. 
\footnote{
We thank A. Doi for these most recent informations. 
}
Therefore, 
we can expected the SN value around $\mathcal{R}_{\rm SN} \sim 10\sim 100$ 
for the nominal value of $S_{\rm E}^{\rm satellite}(\sim 28000$ Jy). 
For the worst value of $S_{\rm E}^{\rm satellite}(\sim 190000$ Jy), the SN value 
is $\mathcal{R}_{\rm SN}<7$, and then useful images cannot be obtained.  
}

We also investigate the effects of the noise for the case of 22 GHz for model B(b)p05 
(see the right panel of Fig \ref{fig:noise}). 
The asymmetric brightness profile can be seen for the cases of 
$\mathcal{R}_{\rm SN}=50$ and 100. From these curves, we can marginally put a constraint 
on the size $\ell_{\rm BH}$ as $\ell_{\rm BH}\sim 30~GM/c^2$ for 
$\mathcal{R}_{\rm SN}\sim 100$ and $\sim 35~GM/c^2$ for $\mathcal{R}_{\rm SN}\sim 50$. 
In these cases, however, we cannot see a deficit of the observed intensity around the black hole.   
In the same way as we did for 43 GHz, 
we estimate the value of $S_{\rm E}^{\rm space}$ satisfying 
the condition of $\mathcal{R}_{\rm SN} \gtrsim 50$. 
By using the values of $S_{\rm E}^{\rm ground}=500$ Jy for the VLBA, $\tau_a=120$ s and 
the same values for other parameters,  
the SEFD at 22 GHz required for the observations satisfying $\mathcal{R}_{\rm SN} \gtrsim 50$ 
becomes $S_{\rm E}^{\rm satellite} \lesssim 2.4\times 10^4$Jy. 
{
At 22 GHz, 
the nominal and the possible worst values of $S_{\rm E}^{\rm satellite}$ 
are 5000 Jy and 8200 Jy for the ASTRO-G satellite. 
\footnote{
We thank A. Doi for these most recent informations. 
}
Even for the worst case, 
we can expected the SN value larger than 100, i.e. $\mathcal{R}_{\rm SN} \gtrsim 100$.  
Therefore, 
}
the observations at 22 GHz by the VSOP-2/ASTRO-G will be able to detect 
the asymmetric brightness profile of the accretion flow around the black hole. 
This means that such observations will give a new evidence of the existence of a black hole 
within the size of $\ell_{\rm BH}\sim 30~GM/c^2$ in M87.

\begin{figure*}
\hspace{-5mm}
\begin{center}
\includegraphics[height=6.0cm,width=16.5cm]{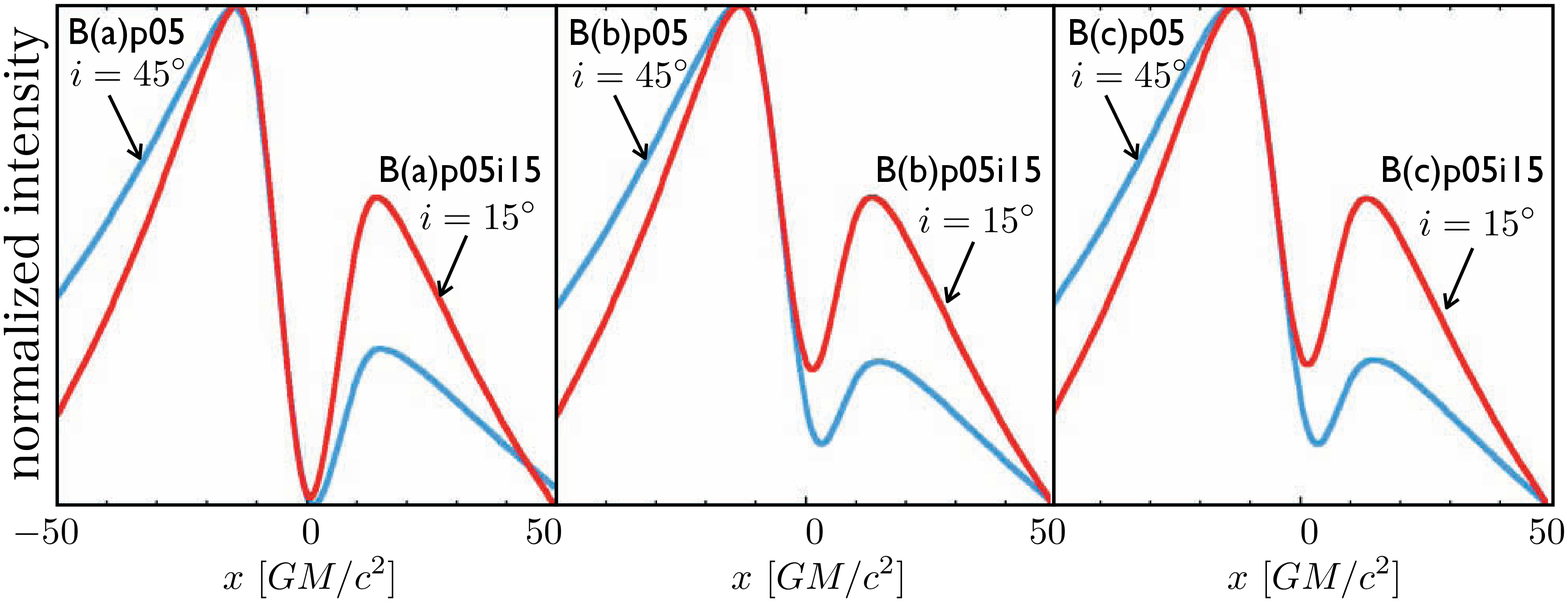}
\end{center}
\caption{\label{fig:va}
Normalized intensity variations along the line passing the black hole center for 
$i=15^\circ$ [models B(a)p05i15, B(a)p05i15 and B(c)p05i15]
and $45^\circ$ [models B(a)p05, B(a)p05 and B(c)p05]. 
Model parameters are given in Table \ref{table:models}. 
}
\end{figure*}

\begin{figure}
\hspace{-5mm}
\begin{center}
\includegraphics[height=8.0cm,width=8.0cm]{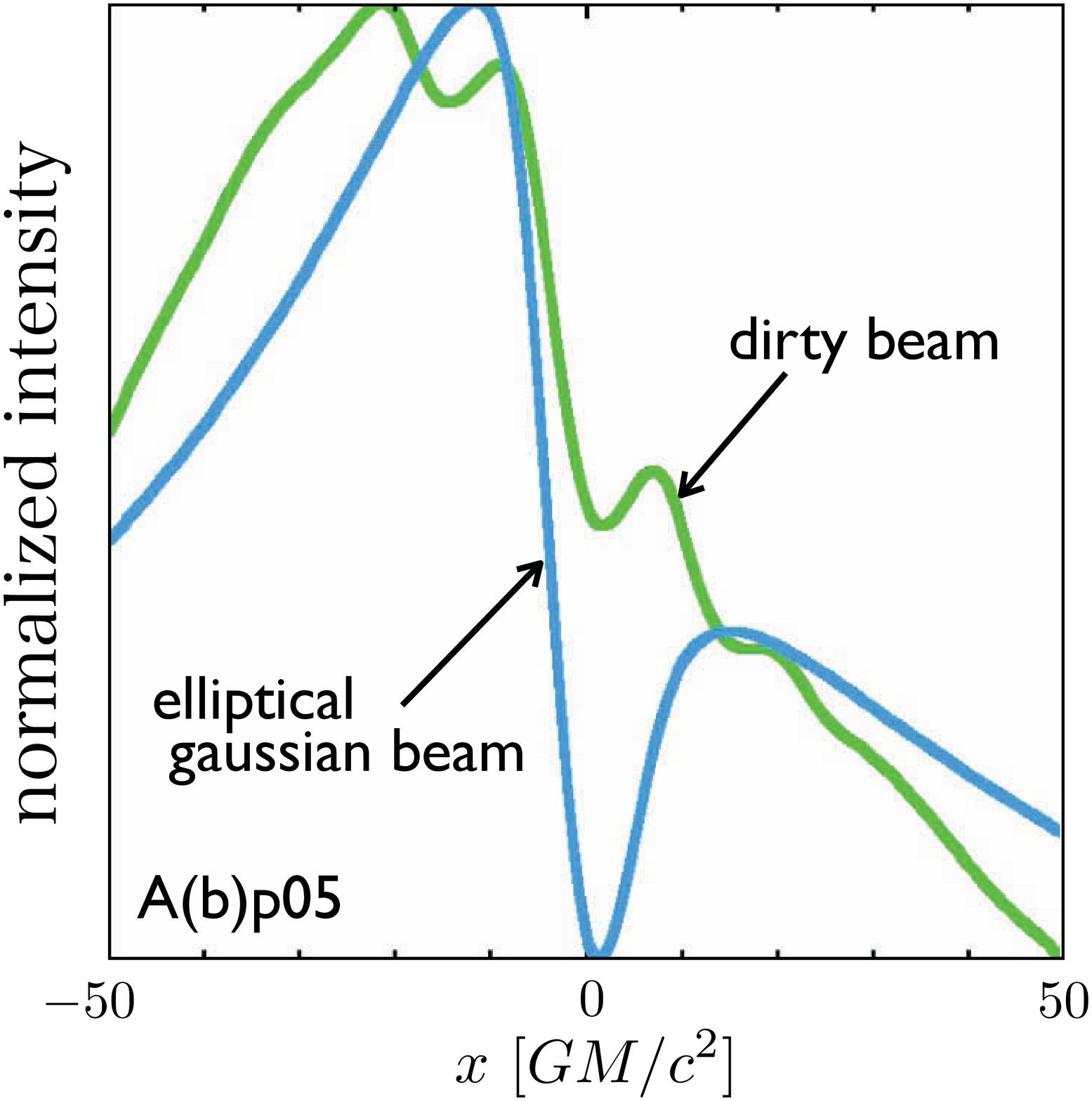}
\end{center}
\caption{\label{fig:db}
Normalized intensity variations along the line passing the black hole center for 
the cases using the dirty beam and the elliptic gaussian beam. 
}
\end{figure}

\section{Discussion}

{
In this study,
we aim at extracting the essential features of the M87 image 
to be observed with a technically feasible space-VLBI satellites.
For the calculation purpose, we have made several assumptions
and simplifications. 
Here, we discuss how realistic our numerical treatments are. 
}

As is well known, bright jets are emerged from the M87 black hole. In this study, 
we implicitly neglected their contribution to the radio emission at 43 GHz and 22 GHz. 
We also neglected the nonthermal electron component. 
These treatments can be justified, since
the outflow component can be negligible at 43 GHz, compared with the 
disk component \citep[][]{j99,dm03,d06},
although it should be included in the other energy bands 
\citep{bl09,l09}. 
Further, the observed energy spectrum can be fitted by the thermal synchrotron 
as was already mentioned in the previous section.
Furthermore, 
since the electromagnetic radiation from the jets
 is likely to be generate by internal shocks, 
which take place at a few tens to a few hundreds of $r_{\rm g}$
apart from the footpoint of the jet; 
that is, the luminous parts of the jets 
should be very distant from the black hole. 
It will be thus easy to discriminate the radio emission 
from the disk from that from the jets. 

In this study, we only consider the sub-Keplerian velocity distribution 
since the flow is very likely to be RIAF, which typically shows sub-Keplerian rotation. 
In the past studies 
\citep[e.g.][]{bl09}, the Keplerian rotation model is considered. Since 
the Doppler boosting effects become much stronger in the case of the Keplarian rotation, 
we expect that 
the asymmetry of the intensity variation shown in Figs \ref{fig:image} and \ref{fig:cs} 
will be more prominent. In the case of the freely falling motion, conversely, 
the asymmetry of the intensity variation will be less pronounced, compared with the sub-Keplerian 
cases considered in this study. 

In this study, we also neglected the effects of a finite disk thickness. 
Since in the case of M87 the viewing angle is not large, i.e. $i\lesssim45^\circ$, 
the effect of the disk thickness is negligible and we have confirmed this by including the effects 
of the thickness for some samples. 
{
Concerning the viewing angle, 
some past studies used more smaller value, such as $i=15^\circ\sim20^\circ$ 
\citep[e.g.][and references therein]{bl09}. 
We have confirmed that the main results in this study will not be changed 
even for such small viewing angles. In Fig \ref{fig:va}, we plot 
the normalized intensity variations along the line passing the black hole center for 
$i=15^\circ$ [models B(a)p05i15, B(a)p05i15 and B(c)p05i15]
and $45^\circ$ [models B(a)p05, B(a)p05 and B(c)p05]. 
The curves for the cases of $i=45^\circ$ are same as those plotted in Figs \ref{fig:image} and 
\ref{fig:cs}. 
From these plots, we can clearly see asymmetric signatures and central dark regions 
for the cases with $i=15^\circ$ as in the cases of $i=45^\circ$. For the cases with $i=15^\circ$, 
the size $\ell_{\rm BH}$ which is the length between the two peaks of the intensity profile is also 
calculated and is summarized in Table \ref{table:models}. The differences in the
intensities of the two peaks in the cases of $i=15^\circ$ are smaller then those in the cases of 
$i=45^\circ$. This is because in the cases of $i=15^\circ$ the Doppler boosting effects become 
smaller than the cases of $i=45^\circ$. For the cases with $i=15^\circ \sim 45^\circ$, 
we expect similar features (e.g. the asymmetric intensity profile, 
the central dark region) as shown in Figs \ref{fig:image} and \ref{fig:cs}. 
}

As shown in the previous section, the effects of the black hole spin can be seen by 
the space-VLBI observations at 43 GHz 
for the case of $r_{\rm in}=r_{\rm ISCO}$ as far as the stationary accretion disk model 
is concerned. 
However, the realistic accretion flow shows significant time variability. 
For the black hole in M87, the Keplerian rotation timescale, $T$, at 
$r_{\rm ISCO}$ is $T\sim 9.2\times 10^4$ s 
$[(r_{\rm ISCO}/r_{\rm g})^{3/2}+a][M/(3.0\times10^9M_\odot)]$ for a distant observer. 
This timescale corresponds to 16 d and 2.1 d for a non-rotating and a maximally 
rotating black hole, respectively, and are both longer than the typical duration of a VLBI 
observation. Thus, the intensity inhomogeneities in the M87 accretion disk and the time 
variable signatures (i.e., flares) will be detected by continuous observation with 
the space-VLBI satellites. 
If such observations are performed, the black hole spin will be measured from the 
variation timescale of the observed flux, and 
the space-VLBI observations 
will be able to confirm the claim by \cite{wlwz08} that the M87 black hole is rapidly rotating. 
Moreover, such inhomogeneity will produce a transient deficit of the observed flux, 
which is similar 
to that around the black hole. However, the former will be time variable while the latter will 
be stable both in time and position. It is thus easy to discriminate one from the other. 
We will investigate the details of these topics in near future.  

We should also discuss what information we can extract if no deficits nor asymmetry were 
detected in the observed intensity profile. 
In such cases, 
it is expected that the velocity patterns should nearly be free-fall as long as the disk is luminous, 
or that  only the jet component can be seen and 
the accretion disk is not luminous due to a low mass accretion rate and/or weak magnetic 
field strength, i.e. radiative efficiency is very small. 

{
In the present study we calculated the smeared images by the convolution of 
the visibility of the theoretical image and 
the Fourier component of the elliptic gaussian beam. 
When actual observational data are obtained, however,
radio observers usually make images by other traditional procedures,
such as the method based on the CLEAN deconvolution algorithm 
(see Sec 11.2 of TMS and references therein). 
We can, in principle and if successful, produce similar images to 
what we calculated here by the CLEAN deconvolution algorithm,
but we should be aware in actual data analyses that
spurious artifacts or some spurious structures are often generated 
in the image during the course of the CLEAN deconvolution algorithm 
(see discussion in TMS). 
When the CLEAN deconvolution algorithm is not successful, therefore,
the effects of the sidelobe pattern cannot be completely removed;
i.e. deconvolution errors remain. 
If this is the case, the calculated images do not reproduce the true images, 
and so 
we cannot correctly extract scientific information from the data. 
In order to see this explicitly, we show in Fig \ref{fig:db}
one extreme case; an image created by the dirty beam for model A(b)p05.
In this figure we can see spurious peaks asymmetric structures
in the case of the dirty beam.  Further we cannot see a central dark region 
around the black hole in the dirty beam case, although it is clearly seen
in the case of elliptical Gaussian beam.
In the analysis of the emission from a luminous region around a black hole, 
it might be difficult to make a correct clean beam map,
since so far there are no good examples reported nor 
successful observations have been made.
This is one example of possible difficulties in the data analysis of 
the observation of a black hole shadow in future. 
In order to reduce such spurious artifacts 
which might be created in images during the process of 
the CLEAN deconvolution method,
other data analysis methods have been proposed; e.g. 
the maximum entropy method (MEM, see, Sec 11.3 of TMS). 
It will be interesting and important as future work
to investigate the validity and the applicability of the data analysis methods,
 such as the CLEAN deconvolution algorithm and the MEM,
in preparation for future interferometric observations of the gas flow 
around black holes. 
}


\section{Conclusions}

To summarize, we investigate the observational feasibility of the central structure  
of the accretion disk, including the black hole shadow, with the space-VLBI observations 
at 22 GHz and 43 GHz by solving the radiative transfer equation in the Kerr space-time 
and adopting the accretion flow model proposed for M87. 
Although in this paper we mainly used the parameters of the ASTRO-G as an example, 
the results obtained in this paper can be applied to other similar space-VLBI missions. 
In this study, we have obtained following results: 
\begin{enumerate}
\item 
Our simulations have demonstrated that the asymmetry 
of the intensity map caused by the disk's rotation can be detected by the space-VLBI 
observations both at 22 GHz and 43GHz. 
That is, the position of the black hole can be determined and this gives the additional 
evidence for the existence of the supermassive black hole in M87.
\item 
In order to achieve these scientific goals and those described below, 
the signal-to-noise ratio $\mathcal{R}_{\rm SN}$ of the observations should be larger 
than at least 10 and desirably 50-100 for the observation at 43 GHz. 
{
If the nominal value of the SEFD of the ASTRO-G satellite at 43 GHz is achieved, 
i.e. $S_{\rm E}^{\rm satellite}\sim 28000$ Jy, we can expect such SN ratios,  
while for the worst case (i.e. $S_{\rm E}^{\rm satellite}\sim 190000$ Jy), such SN 
ratios cannot be expected.   
At 22 GHz, the required SEFD to detect the asymmetric profile is larger than the planned 
value of the VSOP-2/ASTRO-G by several factors even for the worst SEFD value 
(i.e. $S_{\rm E}^{\rm satellite}\sim 8200$ Jy). 
}
The results at 22 GHz will be applied to the observations by the RadioAstron. 
\item 
{ 
From the asymmetric profile, we can put a constraint on the size of the spatial region 
containing the black hole in M87 within 30 $GM/c^2$ 
which is closer to the event horizon than ever before. 
If not, the viscosity 
mechanisms usually proposed by the accretion disk theory and the MHD/kinetic simulations 
should be largely altered or be strongly constrained. 
}
\item
We have shown that 
in the cases that the apparent size of the gravitational radius is 
$r_{\rm g}/D \gtrsim 15\mu$ arcseconds (model A or B in Table 1) 
or that the electron temperature profile is not so steep (i.e., $p\lesssim 0.7$), 
the under-luminous region around the central black hole can also be detected by 
the space-VLBI observations at 43 GHz. 
Such a relatively flat ($p\lesssim 0.7$) 
electron temperature profile is predicted by the accretion disk theory and the three 
dimensional MHD simulations. 
If the under-luminous region is detected, this  gives a very strong evidence 
of the existence of the black hole shadow 
(or black hole silhouette) around the central black hole and 
become the first direct test of the general relativity in the strong gravity region 
around the supermassive black hole. 
\item 
We have also shown in the case of $r_{\rm in}=r_{\rm ISCO}$ that 
the effects of the black hole spin can also be detected by 
the space-VLBI observations at 43 GHz 
If such observation will be performed, the spin value proposed by the past studies  
\citep{wlwz08,l09} will be independently checked and the observation by 
the space-VLBI observations at 43 GHz 
gives more direct evidence of the black hole rotation.  
\end{enumerate}

Finally, since the observational features of the intensity map (including the asymmetry and 
the under-luminous region) depend on the electron temperature profile (see, Fig.\ref{fig:cs}),  
the space-VLBI observations at 43 GHz 
will give the important information of the electron heating 
mechanism which is now one of the biggest unsolved problems 
of relativistic plasmas in curved spacetime.

\acknowledgments
We sincerely thank an anonymous referee for the precious suggestions and comments that 
helped us to improve the original manuscript. 
We would like to thank K. Ohsuga, Y. Kato, Z.-Q. Shen, L. Huang, S. Doeleman, M. Kino, 
M. Takahashi, T. Harada, K. Nakao, H. Ishihara, H. Saida, H. Nagakura, T. Kobayashi, K. Konno     
for valuable discussions and H. Hirabayashi,  M. Inoue, M. Tsuboi, M. Makoto, Y. Murata, 
S. Kameno, Y. Hagiwara, A. Doi, K. Asada, Y. Asaki for useful discussions about  
the observation by radio interferometers and the information about the VSOP-2/ASTRO-G. 
One of the author (RT) would like to thank Dr T. Tamagawa, Profs. K. Makishima and Y. Eriguchi 
for their continuous encouragements. 
This research was partially supported by the Ministry of Education, Culture, Sports, Science 
and Technology (MEXT) by Grant-in-Aid for Japan Society for the Promotion of Science 
(JSPS) Fellows (17010519 R.T.), by the Grant-in-Aid for Scientific Research Fund of the 
Ministry of Education, Culture, Sports, Science and Technology, Japan (Young Scientists(B) 
21740149 R.T.), by Grant-in-Aid of MEXT (19340044, S.M.), and by the Grant-in-Aid for 
the global COE programs on "The Next Generation of Physics, Spun from Diversity and 
Emergence" from MEXT.





\end{document}